\documentclass[preprint]{elsarticle}

\usepackage[utf8]{inputenc}  
\usepackage[T1]{fontenc}     
\usepackage{lmodern}         
\usepackage{microtype}       
\usepackage[scaled]{helvet} 
\usepackage[english]{babel}       
\usepackage[bookmarks=false]{hyperref}            
\usepackage{comment}

\usepackage{amsmath,amsfonts,amssymb}

\usepackage{subfig}

\usepackage[linesnumbered,ruled,vlined]{algorithm2e}
\usepackage{algpseudocode}

\usepackage[T1]{fontenc}
\usepackage{array}
\usepackage{booktabs}

\usepackage{float}

\usepackage{graphicx}
\usepackage{enumitem}

\usepackage{xcolor}
\usepackage{todonotes}

\graphicspath{{fig/}}

\begin{document}

\begin{frontmatter}

\title{A Novel Shortest Path Query Algorithm Based on Optimized Adaptive Topology Structure}

\author[1]{Xiao Fang}
\author[1]{Xuyang Song}
\author[1]{Jiyuan Ma}
\author[2,3]{Guanhua Liu}
\author[3]{Shurong Pang}
\author[4]{Wenbo Zhao}
\author[1,5,6]{Cong Cao}
\author[1,5]{Ling Fan\corref{cor1}}
\cortext[cor1]{Corresponding author}
\ead{fanling@bupt.edu.cn}

\address[1]{School of Electronic Engineering, Beijing University of Posts and Telecommunications, Beijing, China}
\address[2]{School of Transportation, Beijing Jiaotong University, Beijing, China}
\address[3]{Beijing BII-ERG Transportation Technology Co.,Ltd, Beijing, China}
\address[4]{International School, Beijing University of Posts and Telecommunications, Beijing, China}
\address[5]{Beijing Key Laboratory of Space-Ground Interconnection and Convergence, Beijing University of Posts and Telecommunications, Beijing, China}
\address[6]{State Key Laboratory of Information Photonics and Optical Communications, Beijing University of Posts and Telecommunications, Beijing, China}

\begin{abstract}
Urban rail transit is a fundamental component of public transportation, however, commonly station-based path search algorithms often overlook the impact of transfer times on search results, leading to decreased accuracy. To solve this problem, this paper proposes a novel shortest path query algorithm based on adaptive topology optimization called the Adaptive Topology Extension Road Network Structure (ATEN). This algorithm categorizes transfer stations into different types and treats travel time and transfer time equivalently as weights for edges in the topological graph. The proposed algorithm introduces virtual stations to differentiate between pedestrian paths and train paths, eliminating the need for additional operations on transfer stations. The algorithm controls the extent of expansion in the urban rail transit topology, overcoming query errors caused by mishandling of transfer stations in the existing algorithm. Finally, a series of simulation experiments were conducted on Beijing's urban rail transit network to validate both correctness and efficiency of the proposed adaptive topology optimization algorithm. The results demonstrate significant advantages compared to existing similar algorithms.
\end{abstract}
\end{frontmatter}

\section{Introduction} %% {{{

The urban rail transit system possesses distinct advantages in terms of its high capacity, rapid speed, punctuality, minimal environmental impact, and safety and comfort\cite{g}. It has emerged as an indispensable component within urban public transportation systems due to its unique superiority in alleviating congestion in large cities\cite{1}.

As digital technology undergoes rapid advancement, the convenience of people's mobility has significantly increased. The road network structure within urban rail transit systems has become increasingly intricate, characterized by a growing number of transfer stations\cite{2}. Addressing issues pertaining to these transfer stations holds particular significance in route planning and last departure time query algorithms\cite{b}. The escalating number of transfer stations adds complexity to these algorithms, resulting in prolonged computation times\cite{a}. Performing preprocessing on transfer stations prior to route searching and querying the last bus schedule can obviate the need for additional operations on transfer stations, thereby enhancing algorithm conciseness and reducing computation time.

The most commonly employed algorithm for pathfinding in urban rail transit systems is currently the Dijkstra's algorithm\cite{3}.The Dijkstra's algorithm, introduced by the Dutch computer scientist Edsger Dijkstra in 1956, is a renowned method for solving the single-source shortest path problem in a graph. This algorithm efficiently determines the shortest path from a given source node to all other nodes in a weighted directed graph. The fundamental concept of the Dijkstra's algorithm involves iteratively selecting the closest node to the source node and updating distances to its adjacent nodes until either finding the shortest path from source to target or traversing all nodes. Employing a greedy strategy, this algorithm continuously optimizes the path by selecting as an intermediate node the closest unvisited node to the source node and continually updating distance tables\cite{4}.

There are many more algorithms for shortest paths, the Floyd-Warshall algorithm\cite{c}, proposed independently by Robert Floyd and Stephen Warshall in the same year, is a dynamic programming approach for finding the shortest paths between all pairs of nodes in a graph, accommodating both positive and negative edge weights.The A* algorithm\cite{d}, introduced by Hart, Nilsson, and Raphael in 1968, efficiently finds the shortest path in a graph by considering both the actual cost to reach a node and an estimated cost to the goal. 

Several variants have been proposed to improve the efficiency of the algorithm. For instance, Hwan II Kang, Byunghee Lee, and Kabil Kim developed a path planning algorithm by utilizing an enhanced version of Dijkstra's algorithm in conjunction with particle swarm optimization\cite{5}. Debora Di Caprio presented a fuzzy-based Ant Colony Optimization (ACO) algorithm for solving shortest path problems with different types of fuzzy weights\cite{h}. Yijing Chen changed the storage structure, and examined graphing methods, the binary ordering tree algorithm based on the rectangular restricted area reduces the memory storage space\cite{e}. Min Luo,Xiaorong Hou and Jing Yang utilized the Delaunay triangulation to model the surface environment, set up the two-dimensional developable passable channel of the surface and solve the optimal route on this channel\cite{6}.DongKai Fan and Ping Shi improved Dijkstra's algorithm for data storage structures and restricted algorithm search ranges\cite{7}.Yijing Chen analyzed the Dijkstra's algorithm, changed the storage structure, and examined graphing methods, the binary ordering tree algorithm based on the rectangular restricted area reduces the memory storage space\cite{8}. Yuanyuan Hao,Bingfeng Si and Chunliang Zhao proposed a hierarchical network framework based on topology and restoration to search the expected path\cite{hao}.IA Alashoury and MAM Amarif constructed a web application and the typical shortest path between two points of road networks within a specific area was obtained\cite{f}.

In the study of path query in urban rail transit system, Leena Ahmed proposed a sequence-based selection method combined with the great deluge acceptance method achieved the best performance, succeeding in finding improved results in much faster run times over the current best known solutions\cite{i}. Yu Zhou proposed an outer approximation method to linearize the objective to take advantage of existing commercial mix-integer linear programming (MILP) solvers\cite{j}. Rongge Guo suggested a time-dependent bus route planning methodology that explicitly considers path flexibility between nodes to be visited\cite{k}.Lianbo Deng proposed a hierarchical effective path search method made up of a three-layer path generation strategy, which consists of deep search line paths\cite{l}.Fanting Meng proposed an efficient heuristic algorithm is developed to reduce the traffic congestion on an oversaturated urban metro system\cite{m}. Chuan Lin employed the software-defined networking (SDN) technology to improve the scalability of ITS in smart cities and propose a grid-based model to quantify the traffic-congestion probability of the transportation network\cite{n}.Honghao Gao proposed a probabilistic model of the traffic network in the form of a discretetime Markov chain (DTMC) for further computations\cite{o}. Qi Song proposed a dynamic path planning strategy based on fuzzy logic (FL) and improved ant colony optimization (ACO)\cite{p}. Juliana Verga Shirabayashi proposed a algorithm to approache the shortest path problem in colored graphs\cite{q}.

In the context of pathfinding applications in urban rail transit systems, the consideration of transfer time introduces a significant challenge, as the commonly employed Dijkstra's algorithm performs suboptimally when dealing with transfer time. This inadequacy results in inaccuracies in the computation of the shortest path. To solve such problems, Some algorithms based on Dijkstra's algorithm have been proposed. Jimeng Tang, Quanxin Sun and Zhijie Chen solved the problem by expanding the transfer stations.Jimeng Tang, Quanxin Sun and Zhijie Chen also suggested another approach, they changed the label-object from node to edge, and it doesn’t need to expand the network, while taking the time into/outside a station into account\cite{9}. Some characteristics of the related studies are listed in Table ~\ref{Table 1}.

\begin{table}[h]
  \centering
  \caption{Summary of the related study on the shortest path search problem. }
  \tiny 
  \renewcommand{\arraystretch}{1.5}
\begin{tabular}{p{2cm} p{2cm} p{2cm} p{2cm} p{2cm}}
  \toprule
  Methods & Objectives & Mechanism & Advantage & Limitation\\
  \midrule
  Dijkstra + the original topology  & Find the shortest path & Iteratively selecting and updating the nearest node & Simple and easy to implement & Errors occur when considering transfer times \\
  Dijkstra of edge label object + the original topology  & Solve problems and find the shortest path & Treat edges as objects for each iteration & Addresses issues that arise when considering transfer times & There are more objects to traverse \\
  Dijkstra + extended topology  & Solve problems and find the shortest path & Querying the shortest path on the extended topology of the transfer station & Solve some path query error problems & The route found will confuse the driving route and the transfer route \\
  The proposed method  & Solve problems and find the shortest path & Querying the shortest path on the extended topology of the transfer station & Solve problems and find the shortest path & When there are more multi-line transfer stations, the topology will be larger \\
  \bottomrule
  \label{Table 1}
\end{tabular}
\end{table}

In this paper, we propose an effective algorithm to solve the transfer time problem encountered by the traditional Dijkstra's algorithm by preprocessing the transfer stations before querying the shortest path. The optimization algorithm, referred to as adaptive topology optimization algorithm, conducts a more detailed preprocessing of transfer stations before executing the Dijkstra's algorithm for shortest path queries. Specifically, it classifies the transfer stations in conjunction with the actual transfer times, and depending on the scenarios, extends a transfer station into multiple ordinary stations, with the transit times between the stations being the corresponding transfer times, discarding the traditional notion of a transfer station. This eliminates the need to determine whether the current station is a transfer station or not, and the need to retrieve relevant information (e.g., transfer line or transfer direction) to obtain the relevant transfer time.After topology preprocessing, the optimised Dijkstra's algorithm can be used to query the shortest paths without considering any problems related to transfer stations. At the same time, it avoids the problem of wrong query results due to transfer times in traditional algorithms.

In general, the contributions of this paper are summarized as follows: 

$\bullet$ A transfer station classification method is proposed, which categorizes transfer stations based on the transfer time in each direction.

$\bullet$ A new road network structure ATEN is proposed, which is centred on topological extensions for finding the shortest paths between stations in urban rail transit networks.

$\bullet$ Experimental results demonstrate that the proposed adaptive topology optimization algorithm exhibits significant advantages over existing similar algorithms.

The rest of the paper is organised as follows. Section 2 describes the existing applications of Dijkstra's algorithm in urban rail transit. Section 3 demonstrates the topology extension road network structure and describes the process of topology expansion. Section 4 demonstrates the shortest path query after topology expansion. Section 5 verifies the performance of the proposed structure through a series of experiments. Finally, section 6 gives some conclusions.

%% }}} --- Section

%% ---------------------------
%% A section
%% ---------------------------
\section{The application of Dijkstra's algorithm in the context of urban rail transit systems} %% {{{

\subsection{The optimization of Dijkstra's algorithm}

Currently, the commonly employed shortest path query algorithm for rail transit typically utilizes an optimized version of Dijkstra's algorithm\cite{10}. The optimization of Dijkstra's algorithm focuses on two aspects\cite{11}: one is enhancing data storage efficiency and addressing the excessive storage space consumption associated with the classic adjacency matrix by replacing it with an adjacency linked list, thereby effectively reducing overall storage requirements\cite{12}, and the other is to solve the problem of unordered storage of intermediate nodes, heap optimization is used to sort the intermediate nodes in order to reduce the calculation amount of unordered nodes\cite{13}.

\subsection{The procedural steps of Dijkstra's algorithm in urban rail transit system}

The urban rail transit network typically exhibits a sparse graph structure, thus the adjacency linked list is employed for storing topological data, while the binary heap is utilized to optimize and streamline the calculation process in order to minimize computational time.

The procedures are outlined as follows Algorithm 1(~\ref{alg:algorithm1}):

\begin{algorithm}[H]
\SetAlgoNlRelativeSize{-1}
\caption{Dijkstra's Algorithm with Transfer Information}
\KwData{Graph $G$, source node $s$, target node $e$, transfer station information}
\KwResult{Shortest path from $s$ to $e$}
\BlankLine
Initialize array $d[\:]$ with infinity for all nodes to store the distance from the starting point and set $d[s] = 0$\;

Initialize array $p[\:] = \text{null}$\ to store the parent of each node;

Create a set $S$ to keep track of visited nodes\;

\While{$S$ does not contain $e$}{
    Select the node $u$ from $d[\:]$ with the minimum $d[u]$\;
    
    Add $u$ to $S$\;
    
    \ForEach{neighbor $v$ of $u$}{
        \If{is\_transfer[$u$] = 0}{
            \If{$d[u] + w(u, v) < d[v]$}{
                Update $d[v]$ with $d[u] + w(u, v)$\;
                
                Set $p[v] = u$\;
                
                $d.\text{pop}(u)$\;
            }
        }
        \Else{
            \If{$d[u] + w(u, v) + transfer\_time(u,v) < d[v]$}{
                Update $d[v]$ with $d[u] + w(u, v) + transfer\_time(u,v)$\;
                
                Set $p[v] = u$\;
                
                $d.\text{pop}(u)$\;
            }
        }
    }
}
\Return Shortest path from $s$ to $e$;
\label{alg:algorithm1}
\end{algorithm}

In general, Dijkstra's algorithm is a very efficient algorithm to find the shortest path of a single source, which is often used in routing algorithms and network optimization

\subsection{Problems in the application of Dijkstra's algorithm for urban rail transit system}

When applying Dijkstra's algorithm to urban rail transit shortest path query, it is necessary to determine the shortest path to the current station $D_i$ and the corresponding travelling time $min(D_i)$, and further judge whether the station is a transfer station. If it is a transfer station, it is also necessary to consider whether the corresponding transfer time $T_{D_i}$ should be added when determining the travelling time of the station $D_j$ directly adjacent to the transfer stations. For station $D_i (i=1,2,...,n)$, $w(D_i,D_j)$ represents the travelling time from $D_i$ to $D_j$, $T_{D_i}$ represents the transfer time of the line at station $D_i$, The minimum travelling time between $D_i$ and $D_j$ can be expressed as \ref{eq:1}:

\begin{equation}
min(D_j)=min(D_i)+w(D_i,D_j)+T_{D_i}\label{eq:1}
\end{equation}

\begin{equation}
T_{D_i} =
\begin{cases}
    0 & is\_t(D_i) = 0, \\
    t_{Di} & is\_t(D_i) = 1.
\end{cases}
\end{equation}

Here, 

$\bullet$ $t_{Di}$ means the corresponding transfer time of station $D_i$.

$\bullet$ $is\_t(D_i)$ indicates whether station $D_i$ is an transfer station, if $is\_t(D_i)$ equals 1, then station $D_i$ is a transfer station, if $is\_t(D_i)$ equals 0, then station $D_i$ is not a transfer station. 

\begin{figure}[H]
  \centering
  \includegraphics[width=0.3\textwidth]{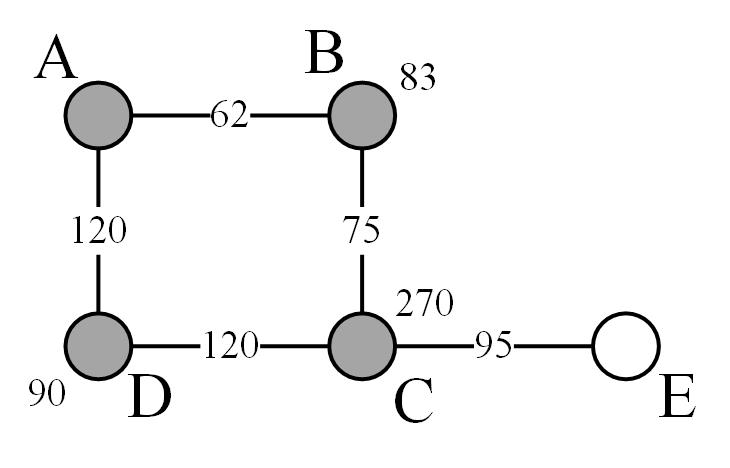}
  \caption{Algorithm Flaw Schematic.}
  \label{fig:Figure1}
\end{figure}

The existing algorithm may encounter errors when the transfer time satisfies certain conditions. The section is shown in figure\ref{fig:Figure1}, $A, B, C, D$ are all transfer stations. The travelling times of the lines and the transfer times of the stations are marked in the figure, and the time unit is seconds ($s$). When using Dijkstra's algorithm to query the shortest path from $A$ to $E$, according to the greedy principle, the algorithm will continuously select the node closest to the source node $A$, and update the distance between the source node and adjacent nodes, until the shortest path between the source node $A$ and the target node $E$ is found.

According to the general Dijkstra's algorithm, the resulting path is $A-B-C-E$, the travelling time $T_{AE}$ is $585s$, as shown in(~\ref{eq:3}), 

\begin{equation}
w(A,E) = w(A,B) + w(B,C) + w(C,E) + T_B + T_C = 585s\label{eq:3}
\end{equation}

However, it is easy to observe from figure\ref{fig:Figure1} that there exists path $A-D-C-E$ in $A-E$, and the travelling time $min(E)$ is $425s$, as shown in(~\ref{eq:4}),

\begin{equation}
min(E) = w(A,D) + w(D,C) + w(C,E) + T_D = 425s\label{eq:4}
\end{equation}

It is obvious that $min(E) < w(A,E)$, that is, there exists a shorter path $A-D-C-E$ between $A$ and $E$ than the path $A-B-C-E$. The reason for the above error is that according to Dijstra's algorithm, the parent node of $C$ is determined to be $B$, so when calculating the shortest path to the target node $E$, the path $A-D-C-E$ is ignored. Since Dijstra's algorithm does not consider transfer time during path finding and all paths passing through a node after determining its parent are determined, the traditional algorithm selects the longer $A-B-C-E$ paths.

\subsection{Existing optimisation programmes}
\subsubsection{Existing topology extension programme}

\begin{figure}[H]
    \centering
    \begin{minipage}[t]{0.3\linewidth}
        \centering
        \includegraphics[width=\linewidth]{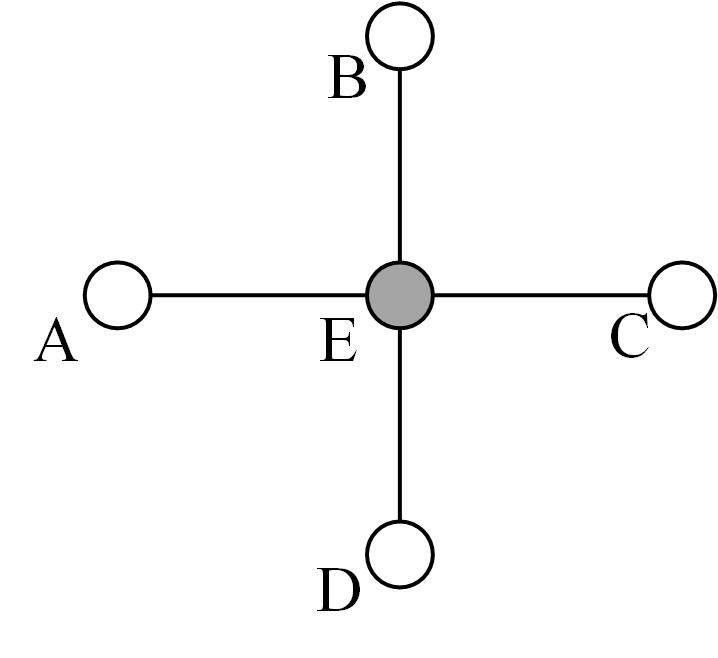}
        \caption{Original topology.}
        \label{fig:Figure2}
    \end{minipage}
    \hspace{1cm}
    \begin{minipage}[t]{0.3\linewidth}
        \centering
        \includegraphics[width=\linewidth]{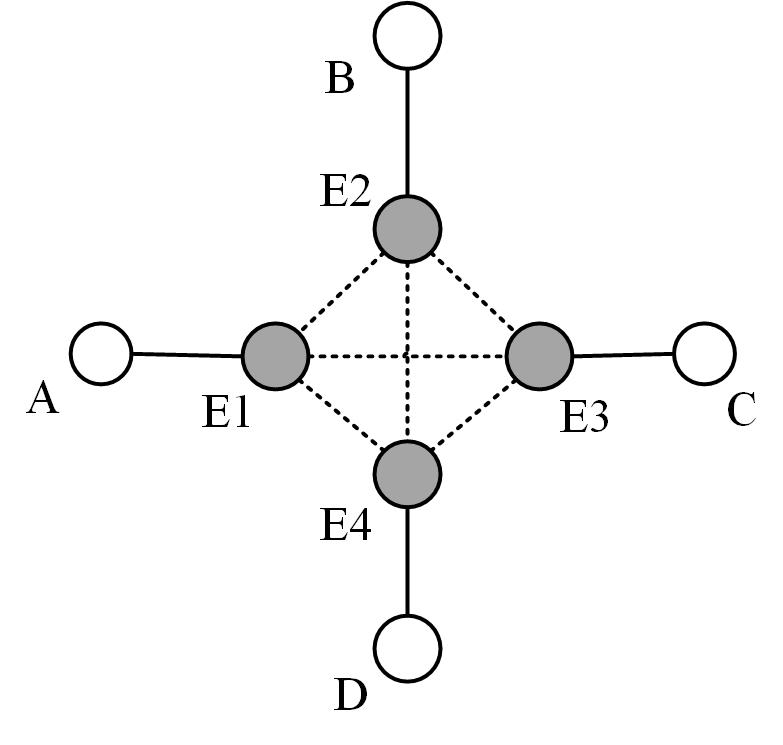}
        \caption{Expanded topology.}
        \label{fig:Figure3}
    \end{minipage}
 \end{figure}

The topology is preprocessed prior to the shortest path query, taking into account the varying transfer times in each direction at the transfer stations.  Specifically, all transfer stations are transformed into four common stations, which are then interconnected by lines.  Furthermore, within this new network configuration, the transfer time in each corresponding direction is assigned as the travelling time between these stations\cite{14}.The original topology is shown in figure\ref{fig:Figure2}, where, E is the transfer stations, $A-E-C$ is line 1, $B-E-D$ is line 2, Take station A, B and E as an example, The travelling time between A and E is $w(A,E)$, The transfer time of station E in direction $A-E-B$ is topology $T_{EAB}$.

The transfer stations E was estended into four stations $E1, E2, E3$, and $E4$, and the addition of lines $E1-E2$, $E2-E3$, $E3-E4$, and $E4-E1$(bi-directional) represents the transfer lines in each direction of the transfer stations E. The addition of lines $E1-E3$ and $E2-E4$(bi-directional) represents the routes that trains directly pass through, as shown in figure\ref{fig:Figure3}.Where, $T_{E1E3} = T_{E3E1} = T_{E2E4} = T_{E4E2} = 0$.

The scheme expands all double-line transfer stations to four transfer stations and adds corresponding lines to solve the shortcomings of Dijkstra's algorithm in handling transfer time. However, there are still potential error situations.

Assuming the departure station is $A$ and the destination station is $B$, if
\begin{equation}
w(E3,E2) < w(E1,E2)
\end{equation}
then,
\begin{equation}
w(A,E1) + w(E1,E3) + w(E3,E2) + w(E2,B) < w(A,E1) + w(E1,E2) + w(E,2B)
\end{equation}
That is, the shortest path from $A$ to $B$ returned by the algorithm is $A-E1-E3-E2-B$, while $E1-E3$ is actually a simulated route for vehicles, which cannot be actually passed by pedestrians, thus causing errors.

In addition, in the actual situation, some transfer stations may have the same transfer time in all directions, so the expansion of transfer stations can be distinguished by whether the transfer time in all directions is the same, which can reduce the storage space of the topology map to a certain expand.

\subsubsection{Dijkstra's algorithm of edge label object}

Dijkstra's algorithm is a classical algorithm for solving single-source shortest path problem. It can effectively find the shortest path from a given source node to all other nodes in the graph. The core idea of Dijkstra's algorithm is to gradually find the shortest path by expanding the set of known shortest paths, until the target node is reached or all nodes are considered.

In Edge-Oriented Dijkstra's algorithm, the algorithm operates on edges. It traverses each edge, selecting the next node on the current shortest path from the source node, and updating the weights of the edges adjacent to the node to find a shorter path. To ensure that the shortest path is always processed, the algorithm uses a priority queue (usually the minimum heap) to maintain the order of candidate paths. Because the transfer stations information is stored in the transfer stations node rather than in the edge, the Edge-Oriented Dijkstra's algorithm does not have the above problems when considering the shortest path problem of transfer time\cite{14}.

However, in the sparse graph of urban rail transit network, the algorithm performance is poor compared with the general Dijkstra's algorithm.

\section{Establish Adaptive Topology Extension Road Network Structure}

The existing topology expansion algorithms divide transfer stations into complete graphs based on transfer routes. Although the topology decomposition process is simple, when two adjacent stations are both transfer stations, the existing algorithms cannot distinguish between vehicle and pedestrian routes, resulting in calculation errors that cannot be actually passed. Our algorithm classifies based on the transfer attributes of adjacent stations. If all adjacent stations are transfer stations, we add virtual stations and connecting lines between virtual nodes and other adjacent nodes. The cross-sectional information is defined as the travel time of vehicles, effectively solving the problem of confusion between vehicle and pedestrian channels.

In order to solve the problem of complex handling of transfer stations in the existing shortest path query schemes, a more detailed preprocessing of transfer stations is carried out before querying the shortest paths using Dijkstra's algorithm. Specifically, the transfer time is incorporated into the category of the actual path in the topological map, and the traditional concept of transfer stations is discarded, so that a transfer station is designed as multiple ordinary stations according to different situations, and the up and down times between these stations are consistent with the corresponding transfer times. In this way, the steps of judging whether the current node is an transfer station and finding relevant information (e.g., change-out line, change-in line, change-direction, etc.) to obtain the corresponding transfer time are eliminated. After completing the topological graph preprocessing, the optimised Dijkstra algorithm can be used directly to query the shortest path without considering the transfer stations and their related times.

\subsection{The process of adaptive topology expansion}

The expansion of transfer stations is mainly related to whether the upward and downward positions of the transfer station on the line to which it belongs are in the same position. The upward direction is the train running along the top of the metro line or in the design direction, and the upward location is the station's boarding position in the upward direction; The concepts of downward direction and downward position are similar to the above. 

The existing topology expansion algorithms divide transfer stations into complete graphs based on transfer routes. Although the topology decomposition process is simple, when two adjacent stations are both transfer stations, the existing algorithms cannot distinguish between vehicle and pedestrian routes, resulting in calculation errors that cannot be actually passed. Our algorithm classifies based on the transfer attributes of adjacent stations. If all adjacent stations are transfer stations, we add virtual nodes and connecting lines between virtual nodes and other adjacent nodes. The cross-sectional information is defined as the travel time of vehicles, effectively solving the problem of confusion between vehicle and pedestrian channels. 

The algorithm is shown below\ref{alg:algorithm2}:                                                                      

\begin{enumerate}
 
% \item[(i)] Input: Importing Rail Transit Data, includes connections between stations, travelling times $T_（OD）$ denotes the travelling time of the train from station $O$ to station $D$, and transfer times  $timeS_{out-in}^{i-j}$ denotes the transfer time from the $out$ direction of line $i$ to the $in$ direction of line $j$ at station $S$, where $out, in \in \{0,1,2\}$, $0$ means no distinction between upward and downward, $1$ means upward and $2$ means downward

\item[(i)] Input: Importing Rail Transit Data, includes connections between stations, travelling times $T_{OD}$ denotes the travelling time of the train from station $O$ to station $D$, and transfer times  $timeS_{out-in}^{i-j}$ denotes the transfer time from the $out$ direction of line $i$ to the $in$ direction of line $j$ at station $S$, where $out, in \in \{0,1,2\}$, $0$ means no distinction between upward and downward, $1$ means upward and $2$ means downward.

\item[(ii)] Traverse all stations, and the current station being traversed is $S$.

\item[(iii)] Check if $S$ is a transfer station: If $S$ is not a transfer station, then revert to step (ii), else if $S$ is a transfer station, then go to step (iv).

\item[(iv)] Assume that the upward and downward neighbours of station $S$ are $S_1^i$ and $S_2^i$ respectively, and delete station $S$ and every line connected to station $S$ on line $i$.

\item[(v)] Determine whether station $S$ is at the same upward and downward positions on line $i$, if the upward and downward positions are in the same location, then go to step (vi); else go to step (vii).

\item[(vi)] Add split station $E_0^i$, connect station $E_0^i$ to station $S_t^i (t=1,2)$ respectively, and the travelling time between station $E_0^i$ and station $S_t^i$ are respectively:

\begin{equation}
T_{E_0^i-S_t^i}^i=T_{S-S_t^i}^i, T_{S_t^i-E_0^i}^i=T_{S_t^i-S}^i
\end{equation}

\item[(vii)] Add split station $E_v^i$, connect station $E_v^i$ to station $S_t^i (t=1,2)$ respectively, and the travelling time between station $E_v^i$ and station $S_t^i$ are respectively:

\begin{equation}
T_{E_V^i-S_t^i}^i=T_{S-S_t^i}^i, T_{S_t^i-E_V^i}^i=T_{S_t^i-S}^i
\end{equation}

Add upward split station $E_1^i$ and downward split station $E_2^i$, connecting station $S_t^i (t=1,2)$ to station $E_k^i (k=t)$; the travelling times between station $S_t^i (t=1,2)$ and station $E_k^i (k=t)$ are respectively:

\begin{equation}
T_{E_k^i-S_t^i}^i=T_{S-S_t^i}^i, T_{S_t^i-E_k^i}^i=T_{S_t^i-S}^i
\end{equation}

\item[(viii)] Station $E_k^i (k=0,1,2)$ is connected to the split station $E_p^j (j\neq i,p=0,1,2)$ of station $S$ on line $j (j\neq i)$, and the travelling times between station $E_k^i$ and station $E_p^j$ are respectively:

\begin{equation}
T_{E_k^i-E_p^j}^i=timeS_{k-p}^{i-j}, T_{E_p^j-E_k^i}^i=timeS_{p-k}^{j-i}
\end{equation}

\item[(IX)] Output: Output adaptive topology extension road network structure.

\end{enumerate}

The flowchart for adaptive topology expansion is shown in Figure\ref{flowchart}:

\begin{figure}[H]
  \centering
  \includegraphics[width=0.8\textwidth]{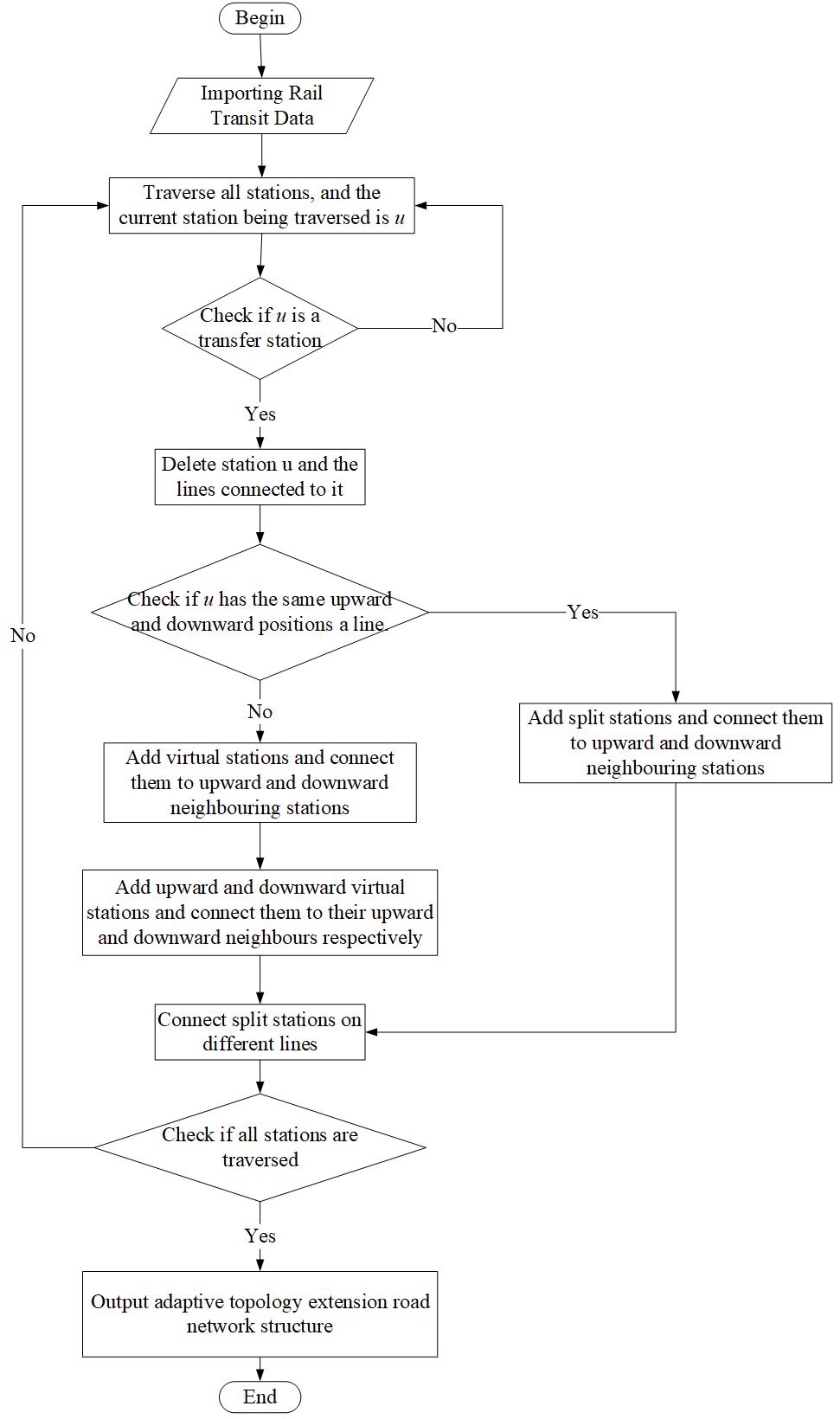}
  \caption{The flowchart for adaptive topology expansion.}
  \label{flowchart}
\end{figure}

\begin{algorithm}[H]
\caption{The Algorithm of Establish Topology Extension Road Network Structure}
\KwData{Rail Transit Data}
\KwResult{Adaptive Topology Extension Road Network Structure}

\While{All stations are not traversed}{
    Traverse all stations
    
    \If{$u$ is a transfer station}{
        Delete station u and the lines connected to it
        
        \If{$u$ has the same upward and downward positions a line}{
           Add split stations and connect them to upward and downward neighbouring stations\;

           Add upward and downward virtual stations and connect them to their upward and downward neighbours respectively
        }
        \Else{
            Add virtual stations and connect them to upward and downward neighbouring stations.
    }
    Connect split stationson different lines
        }
    \Else{
        Continue traversing the station and repeat the above actions
    }

}
\label{alg:algorithm2}
\end{algorithm}

\subsection{Adaptive Topology Extension Road Network Structure.}

Before expansion an transfer station, it is necessary to determine whether the two adjacent stations are transfer stations or not, which can be divided into three cases: neither of the two stations is a transfer station, one of the two stations is a transfer station, and both of the two stations are transfer stations. When both stations are not transfer stations, there is no need to expand the stations and they are directly added to the topology map, while in the other two cases, the stations need to be expanded. 

For transfer stations the classification is as follows, there are four main cases of double-line transfer station and three-line transfer stations:

Case a: The boarding points of the transfer station on each line are at the same place.

Case b: Only one transfer station on the line is located in the same place.

Case c: Only two transfer station on the line is located in the same place.

Case d: The boarding points of the transfer station on each line are not in the same place.

The adaptive topology expansion of various transfer stations is based on the above four cases.

\subsubsection{The transfer stations is adjacent to the non-transfer stations.}

\begin{figure}[h]
  \centering
  \begin{minipage}[b]{0.2\textwidth}
    \centering
    \includegraphics[width=\textwidth]{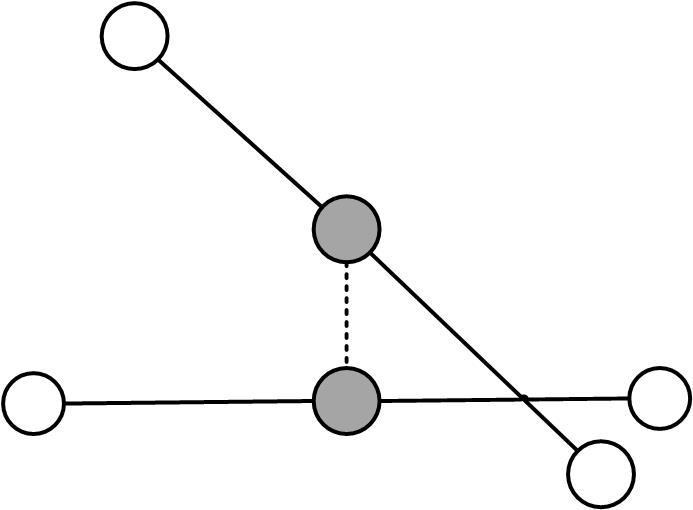}
    \caption*{(a)}
  \end{minipage}
  \begin{minipage}[b]{0.2\textwidth}
    \centering
    \includegraphics[width=\textwidth]{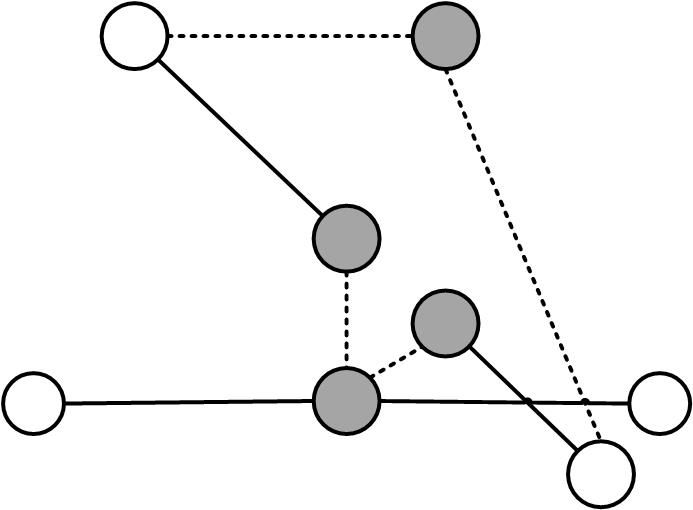}
    \caption*{(b)}
  \end{minipage}
  \begin{minipage}[b]{0.2\textwidth}
    \centering
    \includegraphics[width=\textwidth]{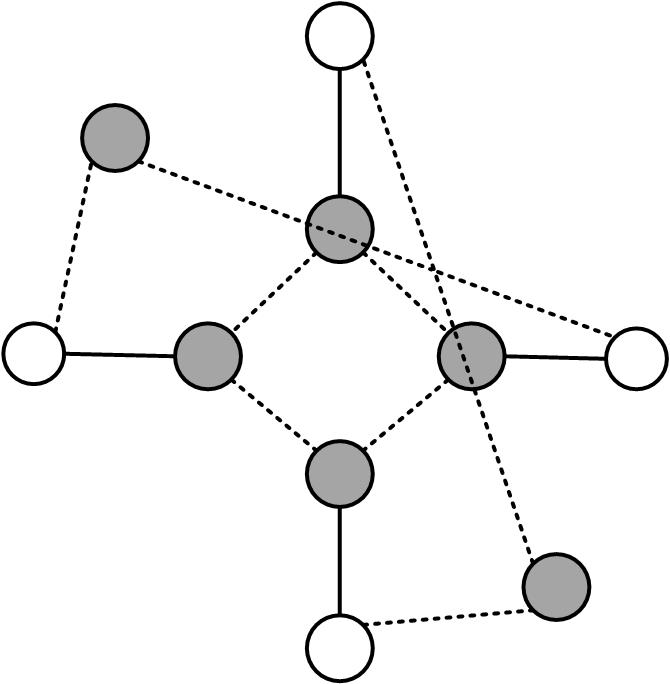}
    \caption*{(c)}
  \end{minipage}
  \begin{minipage}[b]{0.2\textwidth}
    \centering
    \includegraphics[width=\textwidth]{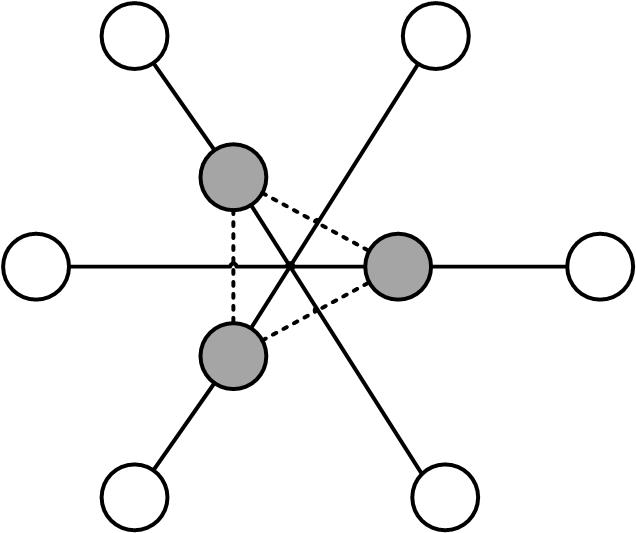}
    \caption*{(d)}
  \end{minipage}
  \begin{minipage}[b]{0.2\textwidth}
    \centering
    \includegraphics[width=\textwidth]{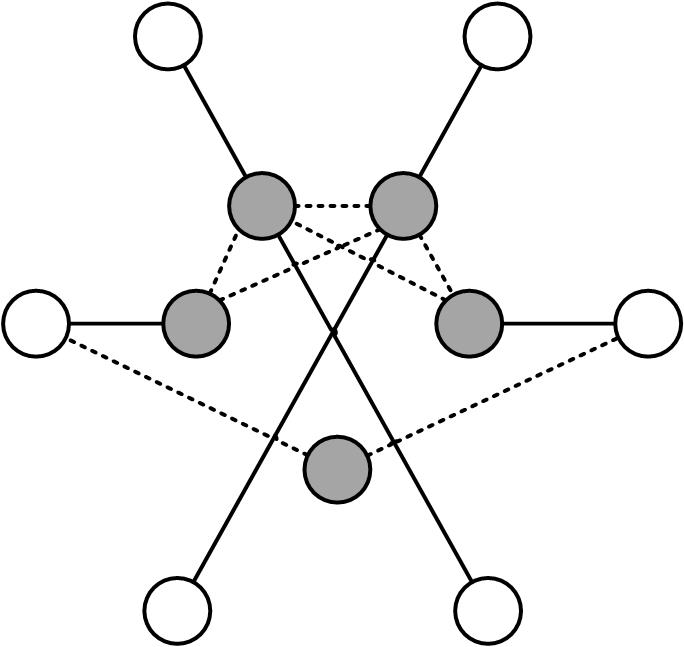}
    \caption*{(e)}
  \end{minipage}
  \begin{minipage}[b]{0.2\textwidth}
    \centering
    \includegraphics[width=\textwidth]{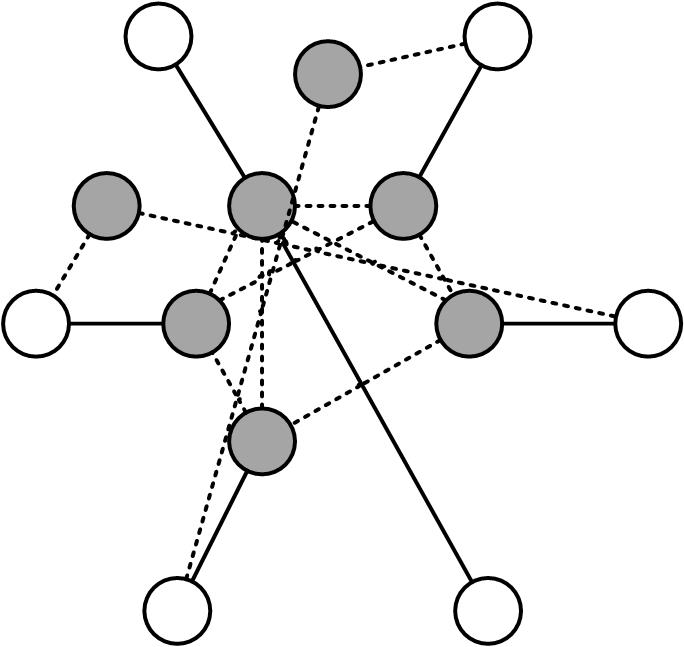}
    \caption*{(f)}
  \end{minipage}
  \begin{minipage}[b]{0.2\textwidth}
    \centering
    \includegraphics[width=\textwidth]{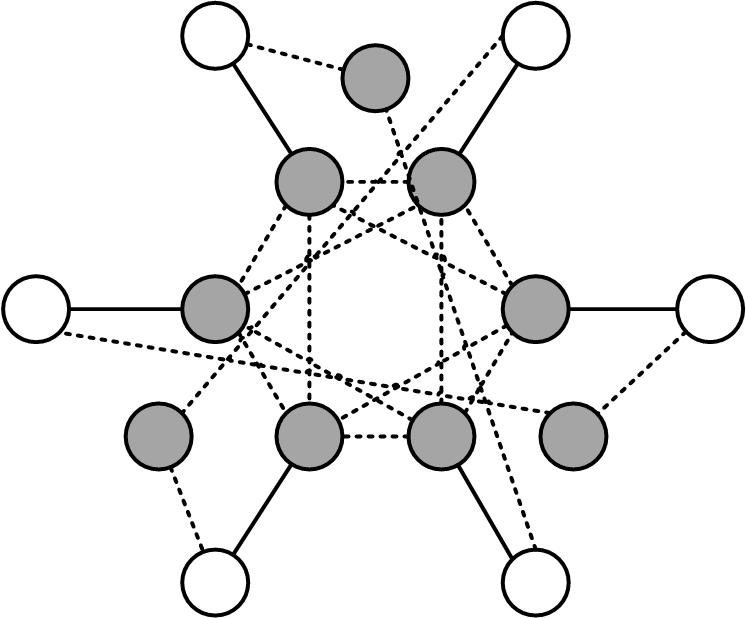}
    \caption*{(g)}
  \end{minipage}
  \caption{Expanded topology of the transfer stations is adjacent to the non-transfer stations.}
  \label{fig:Figure5}
\end{figure}

When only one of the two adjacent stations is an transfer station, simply expand the transfer station and add the corresponding line between the expanded transfer station and the non transfer station. Since the boarding points of up-lines and down-lines at some transfer stations are not at the same location, it may result in different transfer times at the transfer station when the transfer lines are the same but the transfer directions are different, and therefore different lines are needed to indicate the different transfer times. Considering that the transfer lines at the transfer station are generally not more than three, the topology extension road network structures of double-line transfer and three-line transfer are shown in figure\ref{fig:Figure5},where grey nodes indicate stations after the interchange has been expanded, and white nodes are normal stations:

% \begin{figure}[H]
%   \centering

%   \subfigure[]{\includegraphics[width=0.2\textwidth]{erxianquanxingtong.jpg}}
%   \subfigure[]{\includegraphics[width=0.2\textwidth]{erxianyixiangtong.jpg}}
%   \subfigure[]{\includegraphics[width=0.2\textwidth]{erxianquanbutongxiao.jpg}}
%   \subfigure[]{\includegraphics[width=0.2\textwidth]{sanxianquanxiangtong.jpg}}
%   \subfigure[]{\includegraphics[width=0.2\textwidth]{sanxianerxiangtong.jpg}}
%   \subfigure[]{\includegraphics[width=0.2\textwidth]{sanxianyixiangtong.jpg}}
%   \subfigure[]{\includegraphics[width=0.2\textwidth]{sanxianquanbutong.jpg}}

%   \caption{Expanded topology of the transfer stations is adjacent to the non-transfer stations.}
%   \label{fig:Figure5}
% \end{figure}

In figure\ref{fig:Figure5}:

(a) Shows that the double-line transfer station belongs to Case a.

(b) Shows that the double-line transfer station belongs to Case b.

(c) Shows that the double-line transfer station belongs to Case d.

(d) Shows that the three-line transfer station belongs to Case a.

(e) Shows that the three-line transfer station belongs to Case b.

(f) Shows that the three-line transfer station belongs to Case c.

(g) Shows that the three-line transfer station belongs to Case d.

\begin{figure}[H]
  \centering
  \includegraphics[width=0.3\textwidth]{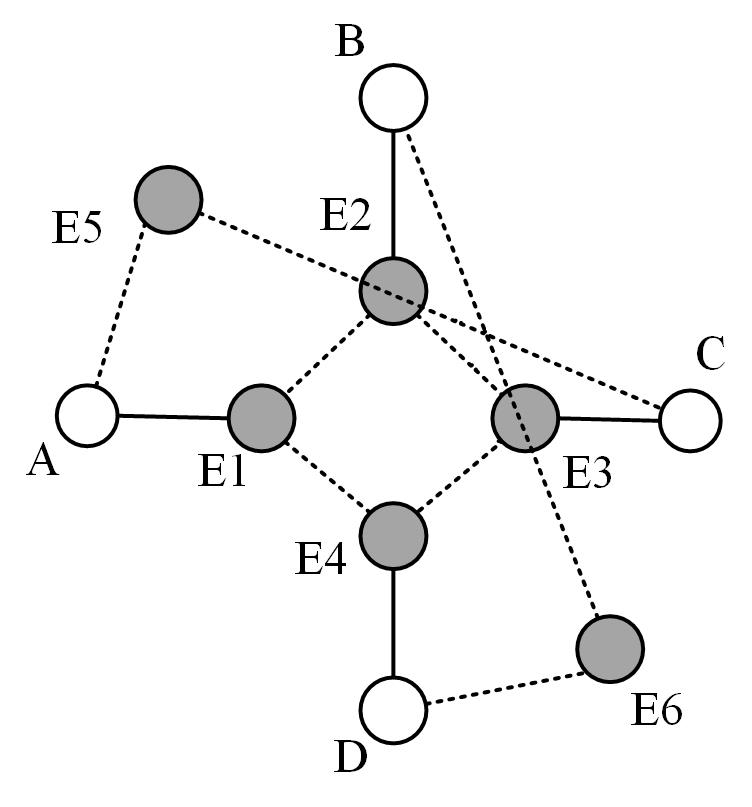}
  \caption{The boarding points for each line are not in the same location.}
  \label{fig:Figure6}
\end{figure}

For example, figure\ref{fig:Figure6} shows a situation in a two-line transfer station where the boarding points for each direction of each line are not at the same location, the boarding places of transfer stations $E$ on line 1 and line 2 are not in the same place, transfer stations $E$ is divided into six stations: $E1, E2, E3, E4, E5$ and $E6$. In order to avoid that the train will pass through the passenger transfer route when querying the shortest path caused by the ordinary segmentation method, this scheme solves the problem by adding $E5$ and $E6$ stations. Add lines $E1-E2, E2-E3, E3-E4, E4-E1$ (bi-directional) to represent the transfer routes in various directions at transfer stations $E$. The $A-E5-C$ (bi-directional) line is introduced to represent the driving paths of $A-E-C$ and $C-E-A$ on line 1 without transfer, and with the same travelling time as the corresponding train on line 1, meanwhile, the $D-E6-B$ (bi-directional) line is added to represent the driving paths of $B-E-D$ and $D-E-B$ on line 2, which require the same travelling time as the corresponding trains on line 2.

\subsubsection{The two adjacent stations are both transfer stations}

\begin{figure}[h]
  \centering
  \begin{minipage}[b]{0.23\textwidth}
    \centering
    \includegraphics[width=\textwidth]{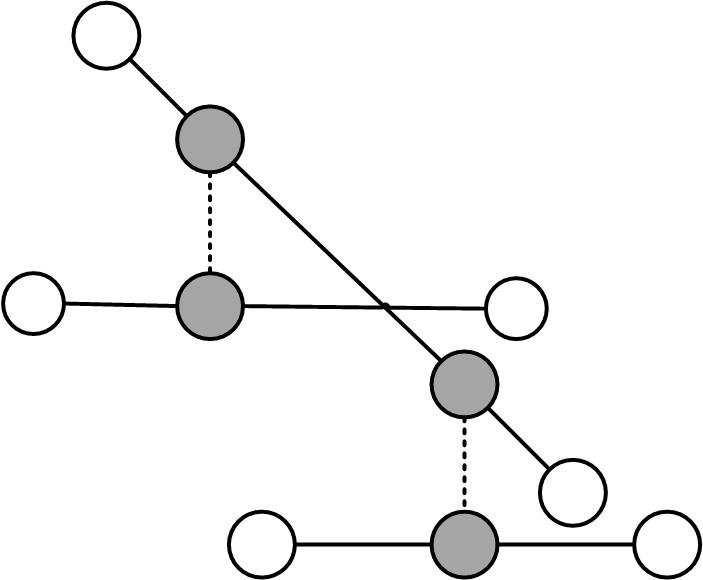}
    \caption*{(a)}
  \end{minipage}
  \begin{minipage}[b]{0.25\textwidth}
    \centering
    \includegraphics[width=\textwidth]{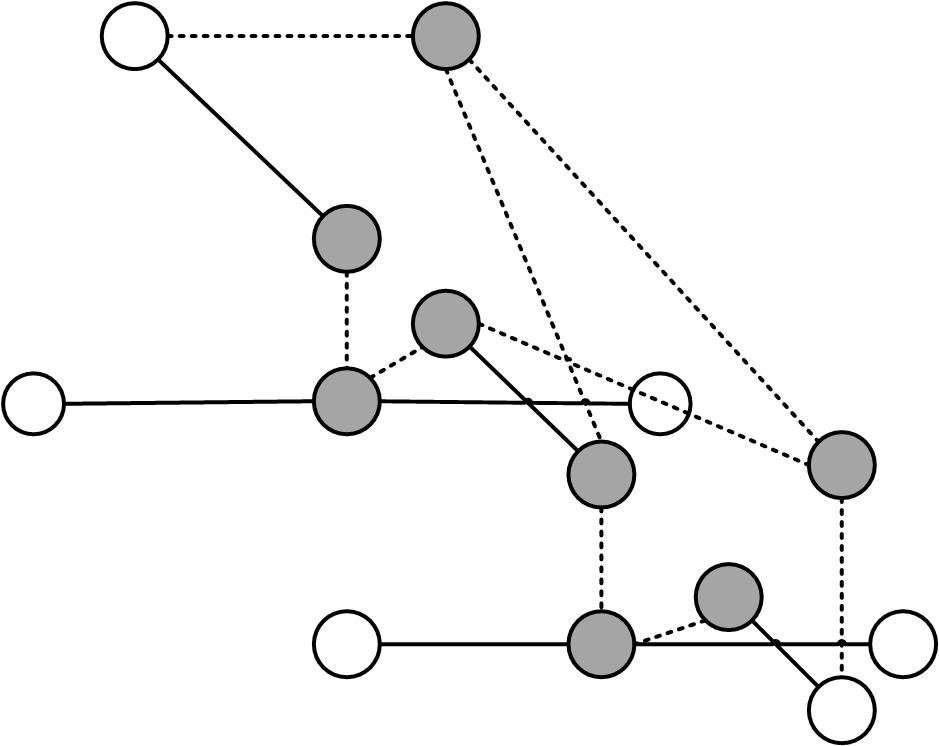}
    \caption*{(b)}
  \end{minipage}
  \begin{minipage}[b]{0.30\textwidth}
    \centering
    \includegraphics[width=\textwidth]{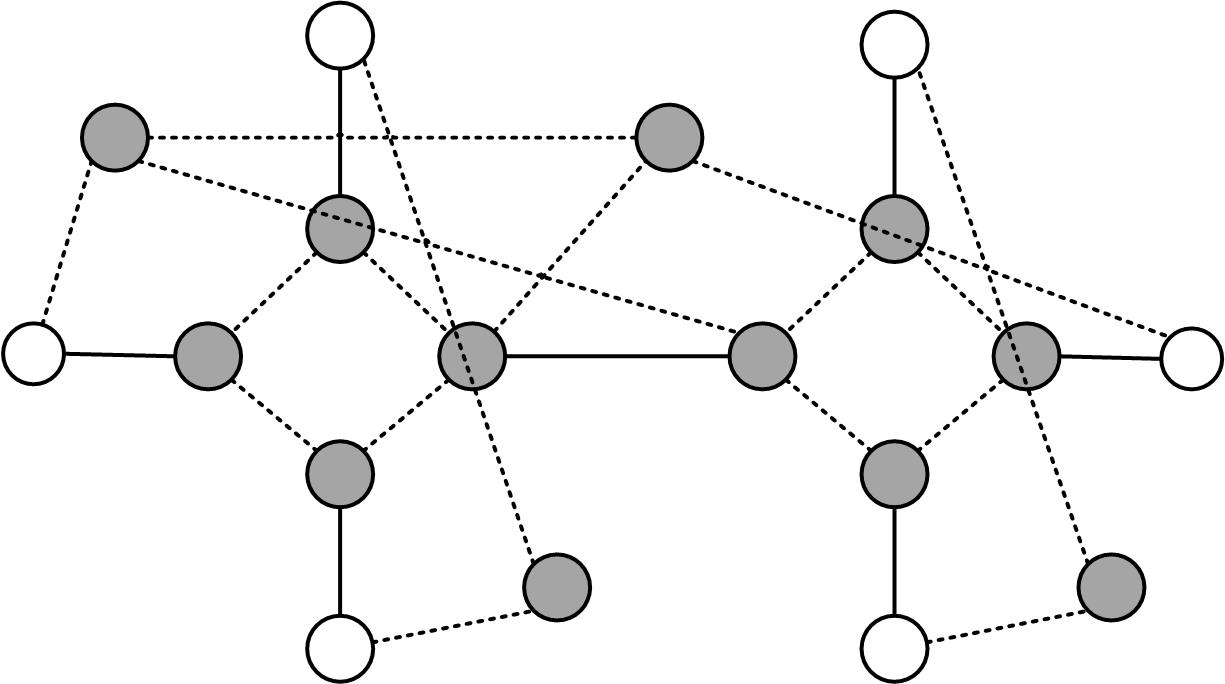}
    \caption*{(c)}
  \end{minipage}
  
  \begin{minipage}[b]{0.25\textwidth}
    \centering
    \includegraphics[width=\textwidth]{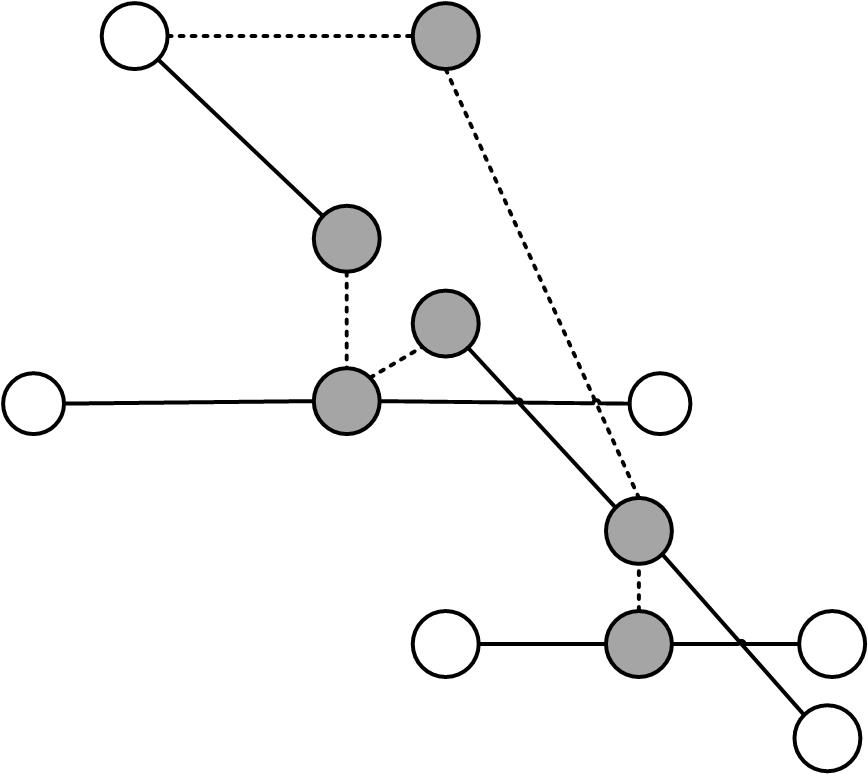}
    \caption*{(d)}
  \end{minipage}
  \begin{minipage}[b]{0.25\textwidth}
    \centering
    \includegraphics[width=\textwidth]{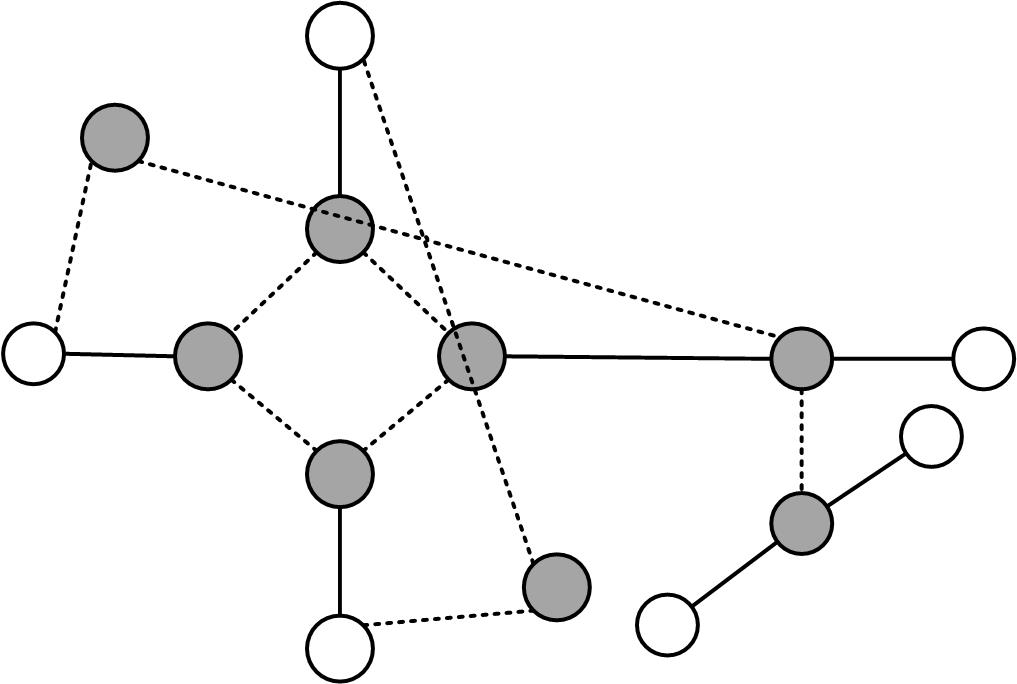}
    \caption*{(e)}
  \end{minipage}
  \begin{minipage}[b]{0.25\textwidth}
    \centering
    \includegraphics[width=\textwidth]{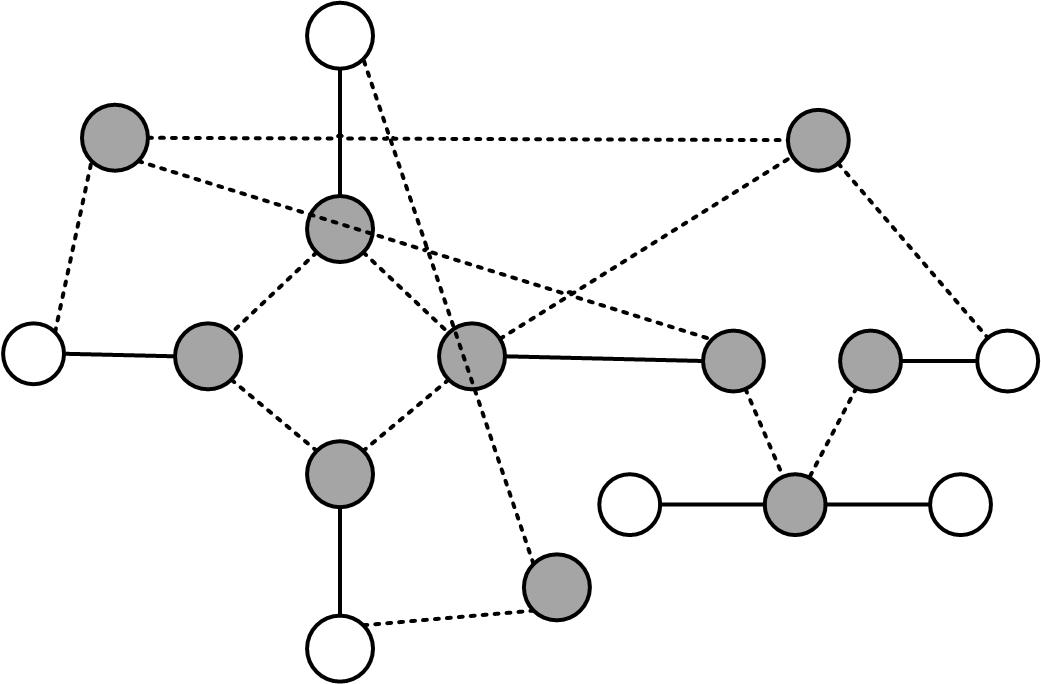}
    \caption*{(f)}
  \end{minipage}
  
  \caption{Expanded topology of two adjacent transfer stations.}
  \label{fig:Figure7}
\end{figure}

In the actual urban rail transit network structure, two adjacent transfer stations mostly adopt the double line transfer mode. There are mainly six options for the topological extension of the network structure of the two adjacent stations, and the topological structure is shown in figure\ref{fig:Figure7}. 

%  \begin{figure}[H]
%   \centering

%   \subfigure[]{\includegraphics[width=0.23\textwidth]{xianglinjunerxianquanxiangtong.jpg}}\label{a}
%   \subfigure[]{\includegraphics[width=0.25\textwidth]{xianglinjunerxianyixiangtong.jpg}}\label{b}
%   \subfigure[]{\includegraphics[width=0.30\textwidth]{xianglinjunerxianquanbutongxiao.jpg}}\label{c}
%   \subfigure[]{\includegraphics[width=0.25\textwidth]{erxianyiquanxiangtong.jpg}}\label{d}
%   \subfigure[]{\includegraphics[width=0.25\textwidth]{erxianqunbutongerxianquanxiangtong.jpg}}\label{e}
%   \subfigure[]{\includegraphics[width=0.25\textwidth]{erxianquanbutongerxianyixiangtong.jpg}}\label{f}

%   \caption{Expanded topology of two adjacent transfer stations.}
%   \label{fig:Figure7}
% \end{figure}

In figure\ref{fig:Figure7}:

(a) Shows that both transfer stations belong to Case a.

(b) Shows that both transfer stations belong to Case b.

(c) Shows that both transfer stations belong to Case d.

(d) Shows that one of the transfer stations belongs to Case a and the other belongs to Case b,

(e) Shows that one of the transfer stations belongs to Case a and the other belongs to Case d,

(f) Shows that one of the transfer stations belongs to Case b and the other belongs to Case d.

\begin{figure}[H]
  \centering
  \includegraphics[width=0.5\textwidth]{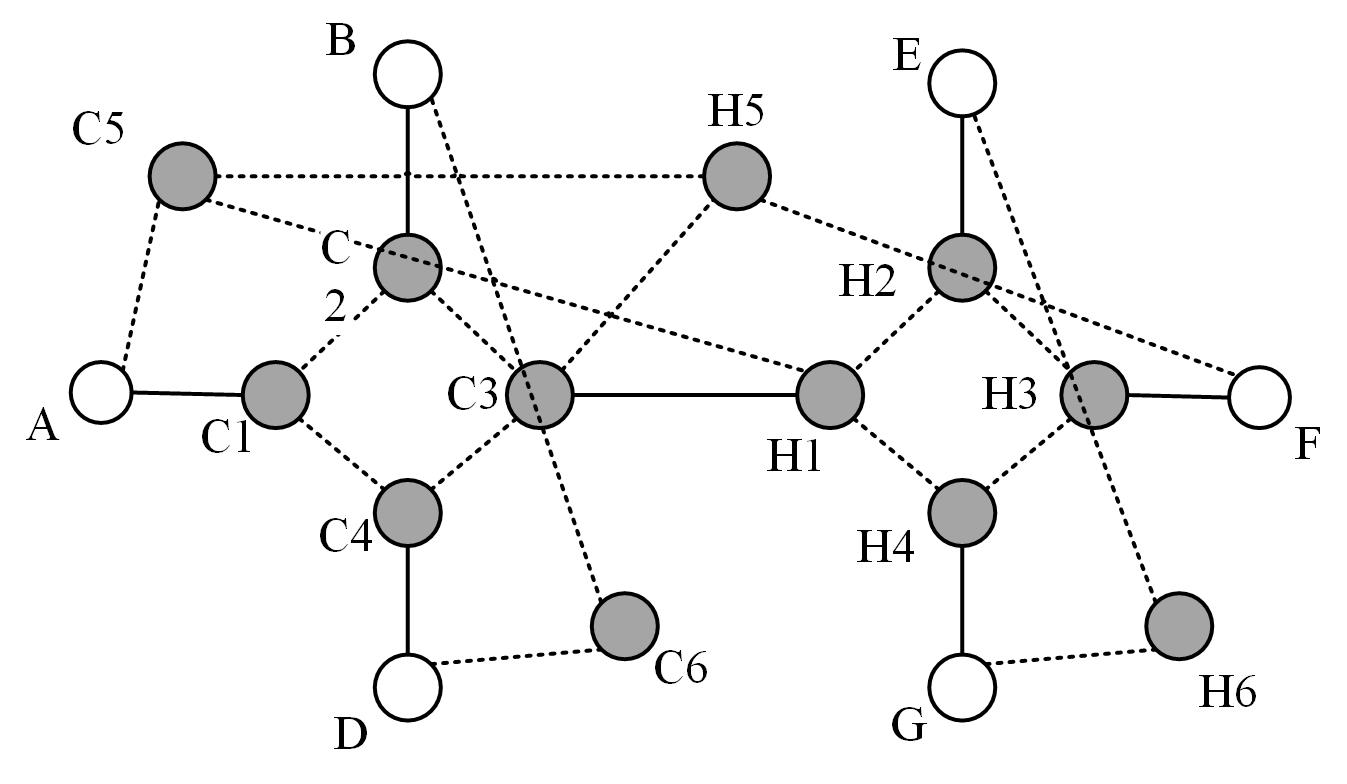}
  \caption{Both adjacent transfer stations are Case 2c.}
  \label{fig:Figure13}
\end{figure}

As shown in figure\ref{fig:Figure13}, both adjacent transfer stations are Case-c, they do not have the same boarding location in all directions, stations $C$ and $H$ on line 1 are transfer stations, the addition of lines $C5-H5$ (bi-directional) to the expansion of the two stations to denote lines on line 1 that pass directly through $A-C-H-F$ (bi-directional) without transfers, and whose travelling time is that between $C$ and $H$ on line 1.

\subsection{Adaptive topology extension for Beijing urban rail transit example}

\begin{figure}[H]
    \centering
    \begin{minipage}[t]{0.3\linewidth}
        \centering
        \includegraphics[width=0.9\linewidth]{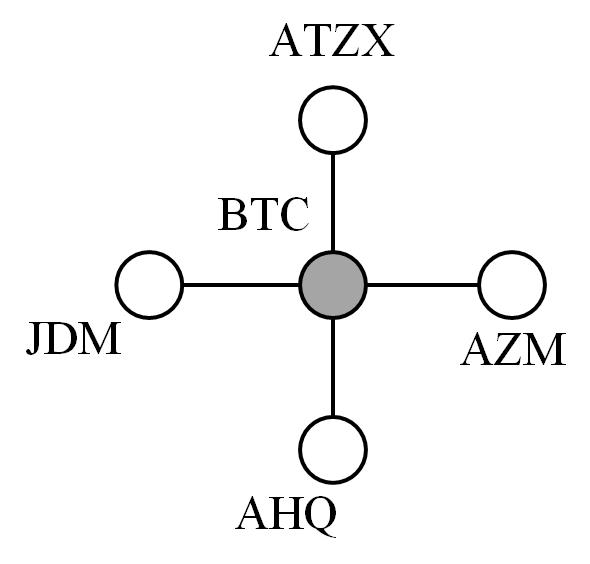}
        \caption{The original topology of $Beitucheng$.}
        \label{fig:Figure14}
    \end{minipage}
    \hspace{1cm}
    \begin{minipage}[t]{0.3\linewidth}
        \centering
        \includegraphics[width=1.2\linewidth]{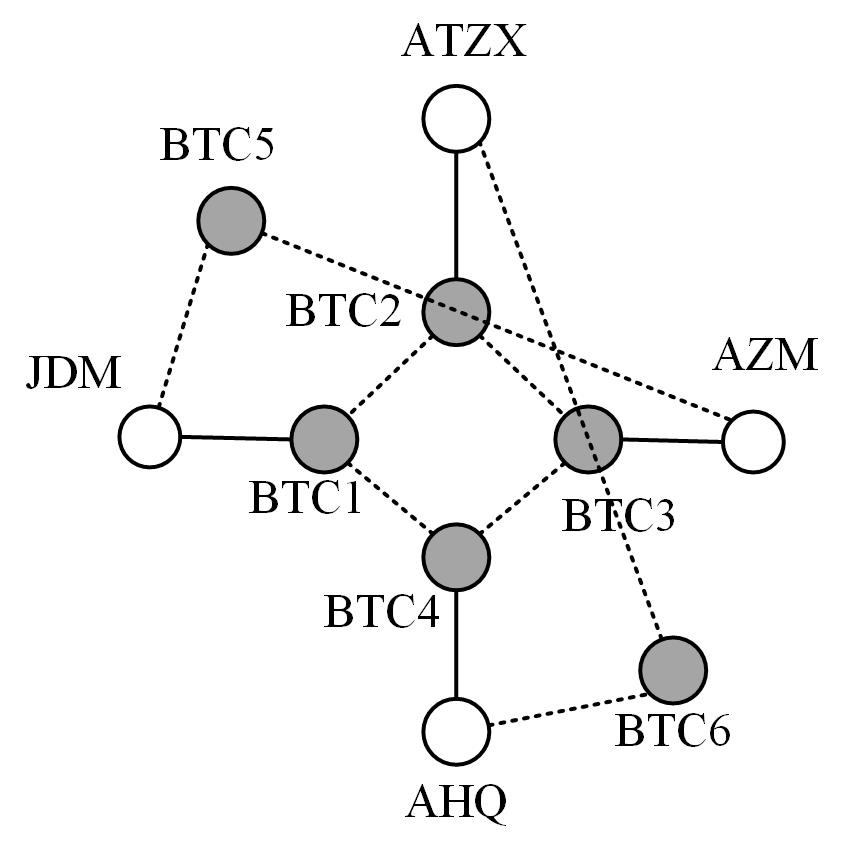}
        \caption{The expanded topology of $Beitucheng$.}
        \label{fig:Figure15}
    \end{minipage}
 \end{figure}

As of March 2023, Beijing rail transit operates 27 lines covering 807 km and has 475 stations, including 81 transfer stations. Take $Beitucheng$(Abbreviated as $BTC$ in the figure) Station as an example, the original topology between the station and its adjacent stations is shown in figure\ref{fig:Figure14}.

The $Beitucheng$ Station belongs to Case a, the boarding points of the transfer stations on each line are at the same place. $Beitucheng$ Station is located on Line 8 and Line 10, and is adjacent to the $Aotizhongxin$(Abbreviated as $ATZX$ in the figure), $Anhuaqiao$(Abbreviated as $AHQ$ in the figure), $Jiandemen$(Abbreviated as $JDM$ in the figure) and $Anzhenmen$(Abbreviated as $AZM$ in the figure) stations. The topology structure of $Beitucheng$ Station and its adjacent stations after topology expansion is shown in figure\ref{fig:Figure15}. $Beitucheng$ Station is divided into six independent stations, namely $Beitucheng 1$, $Beitucheng 2$, $Beitucheng 3$, $Beitucheng 4$, $Beitucheng 5$ and $Beitucheng 6$. Taking $Jiandemen$, $Beitucheng 1$, $Beitucheng 2$ and $Beitucheng 5$ as an example, the travelling time of the line $Jiandemen - Beitucheng 1$ and the line $Jiandemen - Beitucheng 5$ is equivalent to the travelling time of the line $Jiandemen - Beitucheng$ in the original map, and the travelling time of the line $Beitucheng 1 - Beitucheng 2$ is the transfer time required by the transfer from $Jiandemen$ to $Aotizhongxin$ at $Beitucheng$ station.

\section{Shortest Path Query Algorithm Based on Adaptive Topology Optimization}

Once the topology expansion is complete, the ATEN is obtained and then a shortest path query can be performed. Since the concept of transfer stations is abandoned and all stations in the topology are ordinary stations, the shortest path query can be carried out only by using the optimized Dijkstra's algorithm. In the execution of the Dijkstra's algorithm, There is no need to consider whether transfer times need to be added when travelling from the current node to its neighbouring nodes. The algorithm of the shortest path query after the topology extension is as follows\ref{alg:algorithm3}:

\begin{algorithm}[H]
\SetAlgoNlRelativeSize{-1}   
\caption{Shortest path query algorithm based on adaptive topology optimization}
\KwData{$ATEN$, source node $s$, target node $e$}
\KwResult{Shortest path from $s$ to $e$}
\BlankLine
Initialize array $d[\:]$ with infinity for all nodes to store the distance from the starting point and set $d[s] = 0$\;

Initialize array $p[\:] = \text{null}$\ to store the parent of each node;

Create a set $S$ to keep track of visited nodes\;

\While{$S$ does not contain $e$}{
    Select the node $u$ from $d[\:]$ with the minimum $d[u]$\;
    
    Add $u$ to $S$\;
    
    \ForEach{neighbor $v$ of $u$}{
        \If{$d[u] + w(u, v) < d[v]$}{
            Update $d[v]$ with $d[u] + w(u, v)$\;
            
            Set $p[v] = u$\;
            
            $d.\text{pop}(u)$\;
        }

    }
}
\Return Shortest path from $s$ to $e$;
\label{alg:algorithm3}
\end{algorithm}

Compared to Algorithm 1, Algorithm 3 does not have to consider whether the station is a transfer     station or not, and simply query the shortest path directly with an optimised Dijkstra's algorithm.

\section{Simulation Analysis and Application}

To verify the effectiveness and efficiency of the proposed algorithm, a series of experiments were carried out in Beijing rail transit network, in which 380 stations were used, of which 61 were transfer stations and 319 were non-transfer stations. There are three methods to compare with the one presented in this paper:

$\bullet$ method 1: the Dijkstra’s algorithm based on the original topology;

$\bullet$ method 2: the Dijkstra’s algorithm of edge label object based on the original topology;

$\bullet$ method 3: the Dijkstra’s algorithm based on extended topology;

$\bullet$ the proposed method: The shortest path query algorithm based on adaptive topology optimization proposed in this paper.

In this paper, the program corresponding to the above algorithm is developed in Python language. The operating environment of the algorithm is a computer equipped with Intel Core i7 processor (main frequency 2.2GHz), 16GB memory and 1TB hard disk space. The operating system is Windows 10 Home Chinese version, and the programming language is Python 3.9.2. The experimental data set is from the Beijing subway system, which contains the topology of the subway line and the transfer time information.

\subsection{Effectiveness of Dijkstra Based on ATEN}

\begin{figure}[H]
  \centering
  \includegraphics[width=0.6\textwidth]{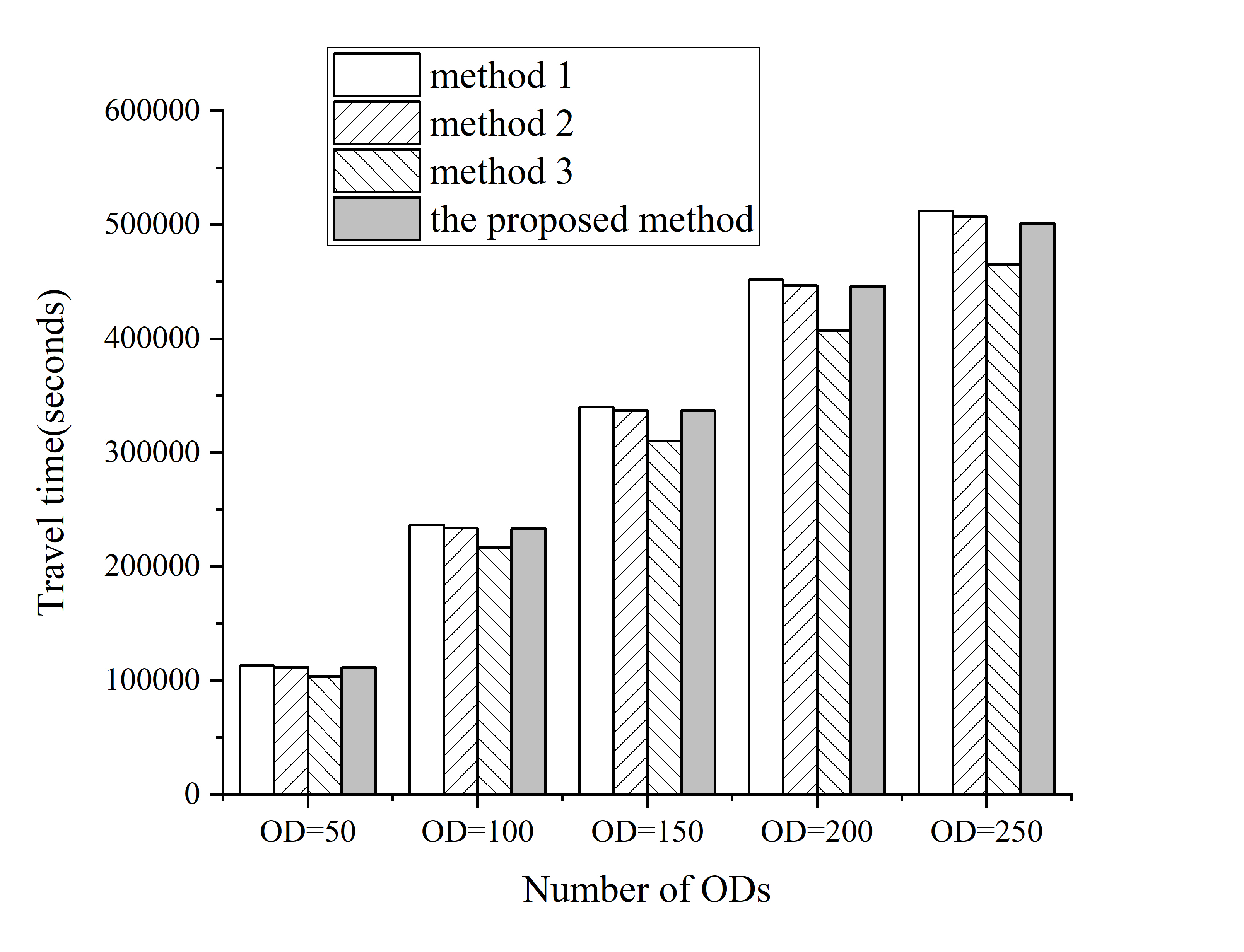}
  \caption{The travelling time of the shortest paths query.}
  \label{fig:Figure16}
\end{figure}

In order to verify the effectiveness of the proposed method, the travelling time of the shortest paths query by the above four methods on the urban rail transit network in Beijing is plotted as follows\ref{fig:Figure16}: 50, 100, 150, 200, and 250 groups of Origin-Destination pairs (ODs) are randomly selected as the inputs of the four methods. In the experiment, each method is guaranteed to use the same ODs as the input in each group of experiments, and each group of ODs contains various combinations of starting stations and terminal stations. The sum of the travelling time calculated by the four algorithms under different ODs groups is respectively counted.

The experimental results shows: the proposed method and method 2 calculation results are basically the same, method 1 calculation results are the largest, method 3 results are the smallest. This shows that when considering the transfer time on the Dijkstra's algorithm, the path calculated by method 1 is longer than the actual shortest path, the proposed method and method 2 calculation results are consistent with the actual shortest path, and method 3 may result in a smaller calculation than the actual shortest path due to the inclusion of paths with a travelling time of $0$ in the expanded stations. The final conclusion is that: the proposed method and method 2 can correctly query the shortest path, and method 1 and method 3 may not be correct in a specific case. That is, The proposed method can query the correct shortest path in all cases.

\subsection{Performance of ATEN on Beijing network}

\begin{figure}[H]
  \centering
  \includegraphics[width=0.6\textwidth]{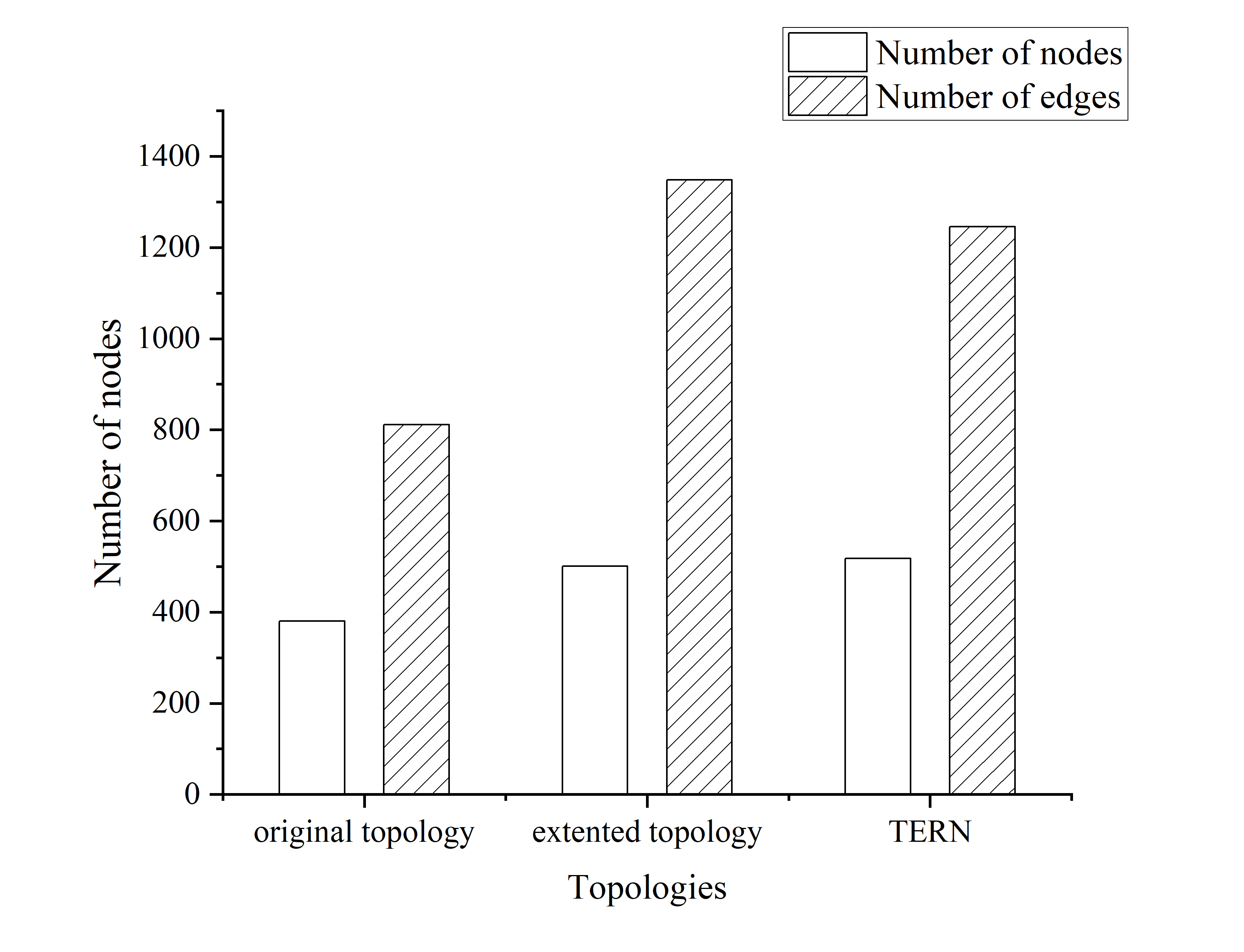}
  \caption{The number of topological nodes and edges.}
  \label{fig:Figure17}
\end{figure}

Methods 3 and 4 both use the extended topology, but the expand and manner of expansion are different. The number of topological nodes and edges used by the two methods are shown below\ref{fig:Figure17}: the original topology has 380 nodes and 811 edges, the topology used by Method 3 has 501 nodes and 1348 edges, and the topology used by the proposed method has 518 stations and 1246 edges. From the perspective of the complexity of the topology, the correct results can be ensured by reasonably expansion the topology in the proposed method.

\begin{figure}[H]
  \centering
  \includegraphics[width=0.6\textwidth]{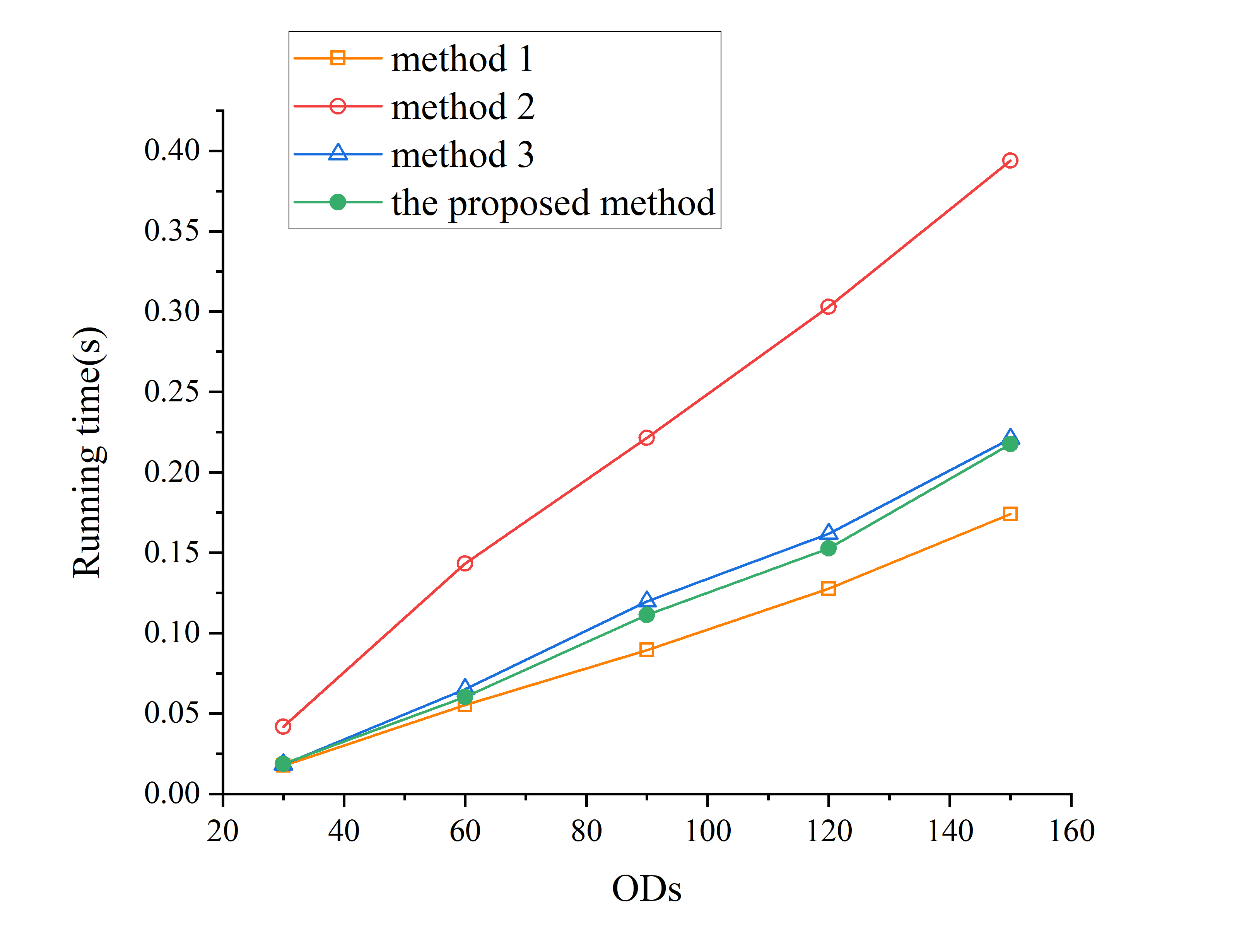}
  \caption{The running time results of each method.}
  \label{fig:Figure18}
\end{figure}

To further verify the performance of the algorithm, the same 30, 60, 90, 120 and 150 groups of ODs were queried using four methods respectively. The running time results of each method are as follows\ref{fig:Figure18}: The running time of the four methods increases with the increase of the number of ODs groups. The total running time of method 1 is the shortest, but the calculation result of this method is not necessarily the shortest path; the running time of method 2 is the longest, but the calculation result is correct; the total running time of method 3 and the proposed method is basically the same, but the calculation result of the proposed method is correct while the calculation result of method 3 is not necessarily correct. It can be seen from the results that the proposed method still has a fast running speed under the premise of ensuring the correct results.

\begin{figure}[H]
  \centering
  \includegraphics[width=0.6\textwidth]{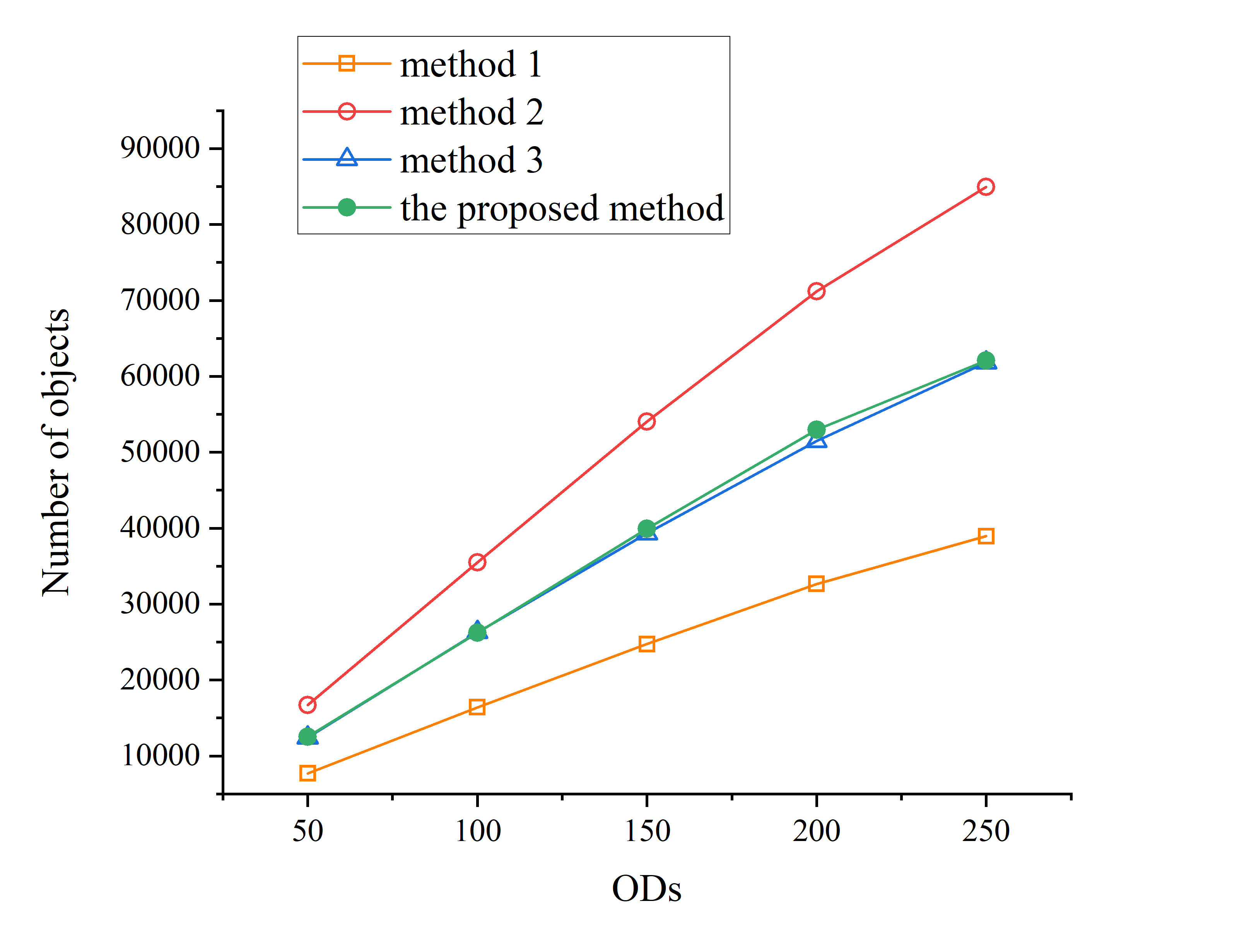}
  \caption{The number of objects traversed}
  \label{fig:Figure19}
\end{figure}

The running time of the algorithm for querying the shortest path is related to the number of objects traversed by the algorithm each time. The following figure shows the sum of the number of objects traversed by four methods in the same 50, 100, 150, 200, and 250 group ODs shortest path query\ref{fig:Figure19}.

As can be seen from the figure, the number of objects traversed by Method 2 is the largest, the number of points traversed by Method 1 is the smallest, and the number of objects traversed by Method 3 and the proposed method is basically the same. The above results are consistent with the travelling time of the four methods.

\section{Conclusions}

The shortest path query algorithm based on adaptive topology optimization proposed in this paper aims to solve the problem of wrong query results when the transfer time is long, and to improve the query efficiency through a series of optimisation measures. In the research process, we have conducted an in-depth analysis of the characteristics of the subway network, and fully tested and verified the algorithm combined with the actual operation situation.

By introducing the topology expansion method, the shortest path query algorithm based on adaptive topology optimization can more accurately evaluate the merits and demerits of candidate paths, avoiding the traditional algorithms which lead to query results because of long transfer time. At the same time, when dealing with large-scale subway network, we adopt the means of optimizing data structure and algorithm, effectively improving the efficiency of query, so that users can get a quick response when querying the shortest subway route.

In future research, we can further optimize the details of the algorithm to improve its applicability in more complex subway networks. In addition, for the emergencies and special situations that may exist in reality, we can also explore more flexible query strategies to cope with different user needs.

In general,the shortest path query algorithm based on adaptive topology optimization proposed in this paper solves the limitations of traditional algorithms and provides a more reliable and efficient navigation scheme for subway travel. We hope that this study can make a certain contribution to the development of urban traffic planning and intelligent navigation system, and provide reference for further research in related fields.

%% }}} --- Section

%% ---------------------------
%% Bibliography and postamblethree
%% ---------------------------
\bibliographystyle{elsarticle-num}

% \section*{References}
\bibliography{main}

\end{document}